\documentstyle[12pt]{article}

\def\ga{\mathrel{\raise.3ex\hbox{$>$\kern-.75em\lower1ex\hbox{$\sim$}}}}
\def\la{\mathrel{\raise.3ex\hbox{$<$\kern-.75em\lower1ex\hbox{$\sim$}}}}
\def\gev{{\rm \, Ge\kern-0.125em V}}
\def\tev{{\rm \, Te\kern-0.125em V}}
\def\beq{\begin{equation}}
\def\eeq{\end{equation}}
\def\ss{\scriptscriptstyle}
\def\mb{m_{\widetilde B}}
\def\msf{m_{\tilde f}}
\def\mtb{\overline{m}_{\ss t}}
\def\mbb{\overline{m}_{\ss b}}
\def\mfb{\overline{m}_{\ss f}}
\def\mf{m_{\ss{f}}}
\def\gt{\gamma_t}
\def\gb{\gamma_b}
\def\gf{\gamma_f}
\def\thm{\theta_\mu}
\def\mgl{m_{\ss \tilde g}}

\begin{document}
\begin{titlepage}
\pagestyle{empty}
\baselineskip=3ex
\rightline{hep-ph/9508413}
\rightline{UMN--TH--1405/95}
\rightline{August 1995}
\vskip.25in
\begin{center}
{\large{\bf
Phases in the MSSM\footnote{To appear in the proceedings of SUSY-95:
International Workshop on Supersymmetry and Unification of
Fundamental Interactions, Palaiseau, France, May 1995}}}
\end{center}
\begin{center}
\vspace*{6.0ex}
{\large Keith A.~Olive
}\\
\vspace*{1.5ex}
{\large \it
{School of Physics and Astronomy,
University of Minnesota, Minneapolis, MN 55455, USA}\\}
\vspace*{4.5ex}
{\bf Abstract}
\end{center}
The effect of CP violating phases in the MSSM on the relic density
of the lightest supersymmetric particle (LSP) is considered.  In particular,
the
upper limits on the LSP mass are relaxed when phases in the MSSM are
allowed to take non-zero values and when the LSP is predominantly a gaugino
(bino).  Previous limits of $\mb \la 250$ GeV for $\Omega h^2 < 0.25$
can be relaxed to $\mb \la 650$ GeV.  The additional
constraints imposed by the neutron and electron electric dipole moments
induced by these phases are also considered. Though there is some restriction
on
the phases, the bino mass may still be as large as $\sim$ 350 GeV and certain
phases can  be arbitrarily large.

\end{titlepage}
\vspace*{4.5ex}
    It is well known that by considering the cosmological relic density
of stable particles, one
can establish mass limits on these particles. In the minimal
supersymmetric standard model
(MSSM), because of the many unknown parameters it is a significantly
complicated task to set
limits on the mass of the lightest particle, the LSP. The
 annihilation cross section will depend
on several parameters which determine the identity of the LSP
as well as parameters such
as scalar masses which mediate LSP annihilations. In addition,
there are (at least) two CP
violating phases in the MSSM which play a role in determining the
LSP relic abundance \cite{fkosi2}.
As will be shown, when the phases are allowed to take their maximal values,
mass limits on the LSP are greatly relaxed. However, constraints from the
neutron electric dipole moment will restrict the degree to which these mass
limits can be relaxed.

The mass and relic density of the LSP depends
 on a relatively large
number of parameters in the MSSM. With some assumptions on
these parameters, their number
can be whittled down to 6: 1) if the gaugino masses are assumed
to obey a GUT relation, they
can be characterized by a single mass parameter, $M_2$; 2) the Higgs mixing
mass, $\mu$; 3) the ratio of Higgs vevs, $\tan \beta$; 4) the soft
(diagonal) sfermion mass parameters, $\msf$ (assumed here to be degenerate
at the weak scale); 5) the soft trilinear terms, $A_f$ (also assumed equal
at the weak scale); and 6) the soft bilinear, $B$.  In general, all but the
sfermion masses, $\msf$, are complex. However not all of the phases
associated with these are parameters are physical \cite{dgh}.
 It is common to rotate away the phase of the
gaugino masses, and to make $B\mu$ real, which ensures that the vacuum
expectation values of the Higgs fields are real. For
simplicity, generation mixing in the sfermion sector will be ignored
though left-right mixing will be included. If furthermore we assume that all
of the
$A_f$'s are equal, and simply label them by $A$, we are left with two
independent phases, the phase of $A$, $\theta_A$, and the phase of $\mu$,
$\theta_\mu$. The phase of $B$ is fixed by the condition that $B\mu$ is
real.

There has been a considerable amount of work concerning phases in the MSSM. For
the most part these phases are ignored because they tend to induce large
electric dipole moments for the neutron \cite{dn,dgh}. If one considers the
direct contribution to the edm from gluino exchange, one finds a dipole moment
\beq
d_n \simeq e {\alpha_s \over \pi} {m_u m_{\tilde g} |\mtb| \over \msf^4} \sin
\gamma_t \sim {\rm few} \times 10^{-23} \sin \gamma_t~{\rm e cm}
\eeq
where the gluino mass, $\mgl$ and $\msf$ as well as $\mtb \equiv A^* +
\mu \cot \beta$ are all taken to be of order 100 GeV. The angle $\gt \equiv
\arg
\mtb$ is  clearly a function of both $\theta_\mu $ and $\theta_A$.
To
suppress the electric dipole moments, either large scalar masses (approaching
$1
\tev$) or small angles (of order $10^{-3}$, when all SUSY masses are of order
100 GeV) are required. For the most part, the community has opted for the
latter, though the possibilities for large phases was recently considered in
\cite{ko}. To reconcile large phases with small electric dipole moments, some
of
the sparticle masses are required to be heavy. In \cite{ko}, it was first
assumed that the SUSY phases were large ($\pi/4$), and by computing the
various contributions to the edm of the neutron from neutralino, gluino,
and chargino exchange (it was the latter that they found to be dominant) either
large sfermion or neutralino masses (or both) were required. For example,
Kizukuri and Oshimo \cite{ko} found that for $M_2
= \mu = 0.5$ TeV, and $\tan \beta = 2$ the chargino contribution to the edm of
the neutron exceeds the experimental upper bound \cite{nexp} of $|d_n| < 1.1
\times 10^{25}$ ecm unless,
$\msf \ga 2$ TeV. When $M_2 = \mu = 0.2$ Tev the limit on the sfermion masses
becomes $\msf \ga 3$ TeV for $\tan \beta = 2$ and $\msf \ga 6$ TeV for
$\tan \beta = 10$.

\vskip 2in

\vskip -.5in
\centerline{{\tenrm Figure 1:Typical contribution to the edm.}}
\vskip .2in

 However, unless
$R$-parity is broken and the LSP is not stable, one requires that
sfermions be heavier than neutralinos, and if they are much heavier, this
would result in an excessive relic density of neutralinos.  Thus one must
determine the relationship between potentially large phases in the MSSM and the
relic density while remaining consistent with experimental bounds
 on the electric
dipole moments.
 Here I will consider only neutralinos as
the LSP. The mass limits will then depend sensitively on the
masses of the scalars. For example when
parameters are chosen so that the LSP is a gaugino (when the
supersymmetry breaking gaugino masses $M_{1,2}$ are taken to be
smaller than the supersymmetric Higgs mixing mass, $\mu$), for a given
LSP mass, the requirement that $\Omega h^2 < 1/4$ places an upper
bound on the sfermion mass ($\Omega$ is the fraction of critical density
and $h$ is the hubble parameter scaled to  100 km/s/Mpc). The reason being that
$\Omega h^2 \propto 1/\langle \sigma v \rangle_A$, and $\langle \sigma v
\rangle_A \propto \msf^{-4}$ is the annihilation cross section.  There is
of course in addition, a dependence on the gaugino mass such that for $m_{\rm
gaugino} < \msf$, a larger gaugino mass implies a larger annihilation cross
section.

 For small gaugino masses, the LSP is a photino, and one obtains a lower limit
on the photino mass
\cite{phot}.  At larger gaugino masses the LSP
is a bino
\cite{osi3} whenever $M_2 < \mu$ and $M_2
\ga 200$ GeV; for smaller $M_2$, the bino is still the LSP for large
enough values of $\mu$.  The bino portion of the $M_2 - \mu$ LSP
parameter plane is attractive, as it offers the largest possibility
for a significant relic density \cite{osi3,gkt,mos}.  The
complementary portion of the parameter plane, with $\mu < M_2$, only
gives a sizable density in a limited region, due to the large
annihilation cross sections to $W^+W^-$ and $ZZ$ and due to
co-annihilations \cite{gs} with the next lightest neutralino (also a
Higgsino in this case), which is nearly degenerate with the LSP
\cite{my}.  For the remainder of the discussion here, I will focus on binos
as the LSP.  In addition to resulting in a sizable relic density, the
analysis is simplified by the fact that in the nearly pure bino region,
the composition and mass of the LSP is not very sensitive to the new
phases. However, as is shown, the relic density, which is
determined primarily by annihilations mediated by sfermion exchange,
is quite sensitive to the phases, $\gamma_t$ and $\gamma_b$.

 Typically, bino annihilation is dominated by the ${\tilde B}
{\tilde B} \rightarrow f {\bar f}$ channel.  When the thermally averaged
cross-section is expanded in powers of $T/\mb$ (see e.g. \cite{swo}), the cross
section can be written as
\beq
\langle \sigma v \rangle_A = a + b T/\mb
\eeq
where
\beq
a \propto {g_1^4 \over 128 \pi} \left( Y_L^2 + Y_R^2 \right)^2 {m_f^2 \over
\Delta_f^2}
\label{a0}
\eeq
\beq
b \sim {g_1^4 \over 128 \pi} \left( Y_L^4 + Y_R^4 \right) {\mb^2 \over
\Delta_f^2}
\eeq
and $\Delta_f^2 = \msf^2 +\mb^2 - m_f^2$. In the above equations, $Y_{L(R)}$
is the hypercharge of $f_{L(R)}$. The expression for $b$ is significantly
more complicated but what is important to note is that $b$ contains a piece
which is proportional to $\mb^2$ whereas $a$ does not. This is a manifestation
of
the p-wave suppression of the annihilation cross-section for majorana
fermions.

 When common sfermion masses are assumed and when sfermion (left-right) mixing
is ignored,  there is an upper limit
$\mb \la$ 250--300 GeV \cite{osi3,gkt} in order to ensure that
$\Omega_{\tilde B} h^2 < 1/4$.  (This upper limit is
somewhat dependent on the value of of the top quark mass. In \cite{osi3,fkmos},
we found that the upper limit was $\sim 250 \gev$ for $m_t \sim 100 \gev$. When
$m_t = 174 \gev$, the upper limit is $260 \gev$. For
$m_t \sim 200 \gev$, this limit is increased to $\mb \la 300 \gev$.
Furthermore, one should be aware that there is an upward correction of about
15\% when three-body final states are included \cite{for} which raises the bino
mass limit to about $350 \gev$. This latter correction would apply to the
limits discussed below though it has not been included.)  As the bino mass is
increased, the sfermion masses, which must be larger than
$\mb$, are also increased, resulting in a smaller annihilation cross
section and thus a higher relic density. At $\mb \simeq 250$ GeV, even
when the sfermion masses are equal to the bino mass, $\Omega h^2 \sim
1/4$. Note that this provides an upper bound on the sfermion
masses as well, since the mass of the lightest sfermion is equal
to the mass of the bino, when the bino mass takes its maximum value.
 One should also note that in the range
20 GeV $\la \mb \la$ 300 GeV, it is always possible to adjust $\msf$ in order
to obtain $\Omega_{\tilde B} h^2 = 1/4$.

 In \cite{fkmos}, it was shown that
the upper limit to the bino mass is sensitive to the level of sfermion mixing.
The general form of the sfermion mass$^2$ matrix is  \cite{er}
\beq
\pmatrix{ M_L^2 + m_f^2 + \cos 2\beta (T_{3f} - Q_f\sin ^2 \theta_W) M_Z^2 &
m_f\,|\overline{m}_{\ss f}| e^{i \gamma_f}
\cr
\noalign{\medskip} m_f\,|\overline{m}_{\ss f}| e^{-i \gamma_f} & M_R^2 + m_f^2
+
\cos 2\beta Q_f\sin ^2
\theta_W M_Z^2
\cr }~
\eeq
where $M_{L(R)}$ are the soft supersymmetry breaking sfermion mass which are
assumed to be generation independent and generation diagonal and hence real.
I also assume
 $M_L = M_R$.
As discussed above, there is a non-trivial phase associated with the
off-diagonal entries, which is denoted by
$\mf(|\overline{m}_{\ss f}| e^{i \gamma_f})$, of the sfermion mass$^2$ matrix,
and
\beq
|\overline{m}_{\ss f}| e^{i \gamma_f} = R_f \mu + A^* = R_f |\mu|e^{i
\theta_\mu}
+ |A|e^{-i \theta_A}
\eeq
where  $R_f = \cot\beta\:(\tan\beta)$ for weak isospin
+1/2 (-1/2) fermions.

\vskip 2in

\begin{center}{{\tenrm Figure 2: Upper and lower limits to the bino mass
as a function of  $A_f$  for $\mu=-3000$ GeV.}}
\end{center}
\vskip .2in

In general, one finds for the s-wave annihilation
cross-section \cite{fkosi2},
\beq
\tilde a_f = { g'^4 \over { 128 \pi } }
  \left| {\Delta_1 ( m_f w^2 - 2 m_{ \widetilde B } w z + m_f z^2 ) +
  \Delta_2 ( m_f x^2 + 2 m_{ \widetilde B } x y + m_f y^2 ) \over
  \Delta_1 \Delta_2} \right|^2
\label{a}
\eeq
where $\Delta_i = m_{\tilde f_i}^2+m_{\widetilde B}^2-m_f^2$ and the factors
$x,y,w,z$ are determined from the bino-fermion-sfermion interaction
Lagrangian,
\begin{eqnarray}
{\cal L}_{f \tilde f \widetilde B } &=& {\textstyle{1\over\sqrt2}}g'
\left( Y_R \bar f P_L \widetilde B \tilde f_R + Y_L \bar f P_R \widetilde B
\tilde f_L \right) + \hbox{h.c.}   \nonumber \\
\noalign{\medskip}
&=& {\textstyle{1\over\sqrt2}}g' \bar f
    \left( \tilde f_1 x P_L + \tilde f_2 w P_L +
\tilde f_1 y P_R - \tilde f_2 z P_R \right) { \widetilde B } + \hbox{h.c.}
\end{eqnarray}
When phases are ignored,  $x = Y_R \sin \theta_f$, $y = Y_L \cos \theta_f$,
$w = Y_R \cos \theta_f$, $z = Y_L \sin \theta_f$. The angle $\theta_f$
relates the sfermion mass eigenstates to the interaction eigenstates by
${\tilde f}_1 = {\tilde f}_L \cos \theta_f + {\tilde f}_R \sin \theta_f$
and ${\tilde f}_2 = - {\tilde f}_L \sin \theta_f + {\tilde f}_L \cos \theta_f$.

When the off diagonal elements of the sfermion mass matrix vanish
 ($|\overline m_f|$
= 0), $\Delta_1 = \Delta_2$ and $xy = wz$ and the pieces proportional to
$\mb$ in (\ref{a}) cancel yielding the suppressed form of the s-wave
cross-section given in (\ref{a0}).
When sfermion mixing is included \cite{fkmos}, the limits, which
now depend on the magnitude of the off-diagonal elements $m_f {\overline m_f}$,
are modified. In Figure 2, the upper and lower limits to the bino mass
is shown as a function of $A_f$, which for fixed
$\mu$ is equivalent to showing the limits as a function
of the off diagonal element
$|\overline m|$.
Because it is the combination $m_f \overline{m}_{\ss f}$ that appears in the
off
diagonal element, sfermion mixing really only affects the stops. Thus when
mixing
is important one of the stop masses is much less than the diagonal
elements $M_{L(R)}$.  In order to ensure that the bino is lighter than
the light stop, the diagonal elements must be pushed up relative to the no
mixing case.  This results in a  reduction in the annihilation cross-section
away
from the no stop mixing  region near $A = 1500$ GeV. The upper limit is
asymmetric
about this point as the couplings of the bino to $\tilde t_1$ and $\tilde t_2$
differ, and the lightest stop is either $\tilde t_1$ or $\tilde t_2$, depending
on
whether or not $A$ is greater than or less than 1500 GeV.
 The lower limit on $\mb$ is easily understood:
when there is a lot of stop mixing and the diagonal elements are large,
the annihilation into channels other than tops, is suppressed.  If $\mb < m_t$,
the annihilation cross-section is too small and $\Omega h^2$ is too large.
Thus we must require the top channel to be available and $\mb > m_t$.
Near the no stop mixing region, we have assumed that $\msf > 74$ GeV
\cite{alitti}. In addition, we are able to obtain $\Omega h^2 = 1/4$ for a wide
range of bino masses (as was the case when sfermion mixing was ignored by
varying
the  the sfermion mass) as well as for a wide range of sfermion masses, by
varying
the magnitude of the off diagonal elements in the sfermion mass matrix.  For
fixed $\mu$ this corresponds to varying $A$.
The upper limit to the bino mass is relaxed considerably
when the phases are allowed to take non-zero values.

As we have just seen that though sfermion mixing has a large effect on
the bino mass limits, the mass limit was not greatly increased. Because
of the mass splitting among the sfermions, ($\Delta_2 \gg \Delta_1$)
there is a reduction in the cross-section which compensates somewhat
for the presence of the term proportional to $\mb$ in $a$.
With non-zero phases, the enhancement in the cross-section is much greater.
Even with $\Delta_2 \simeq \Delta_1$, the terms of interest in (\ref{a})
combine
to
$2 \mb (xy -wz) \ne 0$ which is in fact proportional to $\sin \gamma_f$.
The effect of the phases on the bino mass limits is shown in Figures 3, 4
 as a function of the magnitude of the off-diagonal term in
the top-squark mass$^2$ matrix, $\mtb$, given the conditions: 1)
$\Omega_{\widetilde B} h^2 < 1/4$, $\,$ 2) the lightest sfermion is
heavier than the bino, and 3) the lightest sfermion is heavier than
$74\gev$.  In both figures we have taken $\tan \beta = 2$
and $m_{\rm top} = 174 \gev$. In Figure 3, $|\mu| = 3000$ GeV and in
Figure 4, $|\mu| = 1000$ GeV. The various curves are labeled by the
value of $\gamma_b$ assumed, and in addition, $\mb$ has been maximized
for all allowed values of $\theta_\mu$. The lower limit on $\mb$
assumes $\gamma_b = \pi/2$.  As one can see, when $\gamma_b$ is
allowed to take its maximal value of $\pi/2$, the upper limits are
greatly relaxed to $\mb \la 650$ GeV. With $|\mu|$ and $\gb$ fixed, for a given
value of $\mtb$ and $\thm$ all of the remaining quantities such as $|A|,
\theta_A, \gamma_t,$ and $\mbb$ are determined.

For large $\mtb$, the diagonal mass terms $M_L^2$ must be taken large
to ensure that the mass of the lightest stop is $\ga\mb$.  This drives
up the masses of the other sfermions and suppresses their contribution to the
annihilation.  As $\mtb$ is decreased, $M_L^2$ must drop and the other
sfermions begin to contribute and the upper
bound on $\mb$ is increased.  In particular, annihilation to $\mu$'s and
$\tau$'s becomes important, since
\beq Y_L^2 Y_R^2\,\Big|_{\,\mu,\tau}\; : \;
     Y_L^2 Y_R^2\,\Big|_{\,c, t}\; : \;
     Y_L^2 Y_R^2\,\Big|_{\,s, b} \; = \; 81\: : \: 4 \: : \: 1
\eeq
Decreasing $\gb$ reduces the effect of $\mu$'s and $\tau$'s, and this
can be seen as a decrease in the upper bounds in Figures 3 and 4.
For $\mtb$ sufficiently small, the stops become unmixed, diminishing
somewhat their contribution and slightly decreasing the upper bound
on $\mb$.

\vskip 2in

\begin{center}{{\tenrm Figure 3: Upper limits on the bino mass as a function of
the off- diagonal element $\mtb$ in the top squark mass$^2$ matrix, for various
values of
$ \gamma_b$, the argument of the off-diagonal element of the $T_3 = -1/2$
sfermion mass$^2$ matrix. Also shown is the lower bound (lowest curve) on the
bino mass assuming
$\gamma_b =
\pi/2$. The value of $|\mu|$ was chosen to be $3000 \gev$.}}
\end{center}

\vskip2in

\begin{center}{{\tenrm Figure 4: As in Fig. 3, with $|\mu| = 1000 \gev$.}}
\end{center}

I turn now to the calculation of the electric dipole moments of the
neutron and the electron.  As discussed above, the edm's of the electron and
quarks receive contributions from one-loop diagrams involving the exchange of
sfermions and either neutralinos, charginos, or (for the quarks)
gluinos. In the case of the neutron edm, there are additional
 operators besides the quark electric dipole operator, $O_\gamma = {1 \over 4}
{\bar q} \sigma_{\mu \nu} q {\tilde F}^{\mu \nu}$
\cite{dn} which contribute. They are the gluonic operator
$O_G = - {1 \over 6} f^{abc}G_aG_b{\tilde G_c}$ \cite{w} and the quark color
dipole operator, $O_q = {1 \over 4}
{\bar q} \sigma_{\mu \nu} q T^a {\tilde G_a}^{\mu \nu}$ \cite{ads}. The gluonic
operator is the smallest \cite{other,ads} when all mass scales are taken to
be equal. These three
operators are conveniently compared to one another in \cite{aln} and
 relative to the
gluino exchange contribution to the $O_\gamma$ operator, it was found that
$O_\gamma : O_q : O_g = 21 : 4.5: 1$.

Because of the reduced importance of the additional operators contributing to
the neutron edm, it is sufficient to only include the three contributions to
the
quark electric dipole moment. The necessary
$C\!P$ violation in these contributions comes from either
$\gf$ in the sfermion mass matrices or $\theta_\mu$ in the neutralino and
chargino mass matrices.  Full expressions for the chargino, neutralino and
gluino exchange contributions are found in \cite{ko}. The dependencies of the
various contributions on the CP violating phases can be neatly summarized: for
the
chargino contribution
\beq
d_f^C \sim \sin\thm\;,
\eeq
with essentially no dependence on $\gf$; whilst for  the gluino contribution
\beq
d_f^G \sim \mfb \sin\gf\;,
\eeq
independent of $\theta_\mu$, and the neutralino contribution has pieces that
depend on both $\sin\thm$ and $\mfb\sin\gf$. All three contributions
can be important (including the neutralino contribution, in the case
of the electron edm), and depending on $\sin\thm$ and $\sin\gf$, they
can come in with either the same or opposite signs.  In particular,
${\rm sign}[d_f^C/d_f^G] = {\rm sign}[\sin\thm/\sin\gf]$.  For the
mass ranges we consider, the dipole moments fall as the sfermion
masses are increased, and sfermion masses in the $\tev$ range can
bring these contributions to the
neutron and electron electric dipole moments below the
experimental bounds of $|d_n| < 1.1 \times10^{-25}e\:{\rm cm}$ \cite{nexp} and
$|d_e| < 1.9\times10^{-26}e\:{\rm cm}$ \cite{eexp},
even for large values of the $C\!P$ violating
phases \cite{ko}.  However, these large sfermion masses are inconsistent with
the cosmological bounds mentioned above, where sfermion masses must be
relatively close to the bino mass in order to keep the relic density in check.

To determine the allowed parameter ranges
\cite{fkosi2}, we first fix the value of $\gb$ and
take
$|\mu| = 3000\gev$.  Then for several values of $\mtb$ between 0 and 1500
$\gev$, we determine the upper bound on $\mb$, as a function of
$\thm$.  As we vary $\thm$ across its full range, $\mbb$ and $\gt$
change, and this affects the annihilation rate
 and consequently the bound on $\mb$.  Taking
$\mb$ at its maximum value allows us to take $M_L^2$ as large as
possible; although the electric dipole moments depend on $\mb$
as well, the dependence on
$M_L^2$ is sufficiently strong that the edm's take their minimum
values for the maximum values of $\mb$ and $M_L^2$.
The quark and electron edm's can then be computed as a function of $\thm$ and
$\mtb$,  and one can use the nonrelativistic quark model
to relate the neutron edm to the up and down-quark edm's via
\beq
d_n = {1\over3}(4d_d - d_u).
\eeq
If there is no region of the $\thm$--\,$\mtb$ parameter space which satisfies
both the neutron and electron edm bounds, we  can decrease $\gb$ and repeat
the procedure.  In practice, the bound on the neutron edm
the more difficult of the two to satisfy, and every region of the
parameter space we show which produces an acceptable neutron edm also
produces a sufficiently small electron edm.

For the large value of $|\mu| = 3000\gev$, the largest
contribution to the neutron edm comes either from gluino exchange
(for the more negative values of $\thm$) or chargino exchange (for the
more positive values of $\thm$),
and the value for $|d_n|$ is too large unless $\gb$ takes a relatively
small value.  In particular, non-negligible experimentally
acceptable regions of the parameter space are found only for $\gb \la \pi/25$.
In Figure 5, a contour plot of the neutron edm as a function
of $\thm$ and $\mtb$ for $\gb = \pi/40$ is shown
\cite{fkosi2}.  The shaded regions demarcate
the range of $\thm$ for this choice of $\gb$.  Much of this range
produces a sufficiently small $|d_n|$.  As we increase $\mtb$, the
$\tilde d$ and $\tilde u$ masses become large and $|d_n|$ falls.  As we
move to values of $\mtb$ greater than $\sim 1500\gev$, we begin to
require a significant tuning of $M_L^2$ to produce
$\Omega_{\widetilde B} h^2 < 1/4$.

\vskip 2in

\begin{center}{{\tenrm Figure 5: Contours of the neutron electric
dipole moment, $d_n$, in the
$\thm$ -- $\mtb$ plane in units of $10^{-25}e\:{\rm cm}$. The value of $|\mu|$
was chosen to be $3000 \gev$ and $\gamma_b = \pi/40$. The shaded region
corresponds to values of
$\thm$ and $\mtb$ which are not allowed algebraically for this value of $\mu$
and
$\gb$.}}
\end{center}

This procedure can be repeated for $|\mu| = 1000\gev$ and the result
is shown in Figure 6 \cite{fkosi2}.  For lower
$|\mu|$, the chargino exchange contribution is enhanced relative to the gluino
exchange contribution.  In this case, one can allow larger values of $\gamma_b$
up to about $\pi/6$. For the same range in $\mtb$, $\theta_\mu$ can
takes values from -0.3 to 0.2.  For either value of $\mu$, the angles
$\gamma_t$ and $\theta_A$ are unconstrained.

In summary, we have found that $C\!P$ violating phases in the MSSM can
significantly affect the cosmological upper bound on the mass of an
LSP bino.  In particular, taking the maximal value $\pi/2$ for the
phase $\gb$ of the off-diagonal component of the $T_3 = -1/2$ sfermion
mass matrices pushes the upper bound on $\mb$ up from $\sim 250\gev$
to $\sim 650\gev$.  When we additionally consider constraints on
neutron and electron electric dipole moments, we find the upper bound
on $\mb$ is reduced to $\sim 350\gev$.  Various combinations of the
$C\!P$ violating phases are constrained as well: $|\thm|\la 0.3$ and
$|\gb|\la\pi/6$ for $|\mu| \ga 1000 \gev$, while $\gt$ and $\theta_A$
are essentially unconstrained.
We note that although the bounds on $\thm$ and $\gb$ are small, they
are much larger than the values of order $10^{-3}$ typically considered.

\vskip 2in

\begin{center}{{\tenrm Figure 6: As in Figure 5, with $|\mu|$
chosen to be $1000 \gev$ and $\gamma_b = \pi/8$.}}
\end{center}

\vskip 1in
\vbox{
\noindent{ {\bf Acknowledgments} } \\
\noindent  The work described here was done in collaboration with T. Falk and
M. Srednicki.  This work was supported in part by DOE grant
DE--FG02--94ER--40823.}


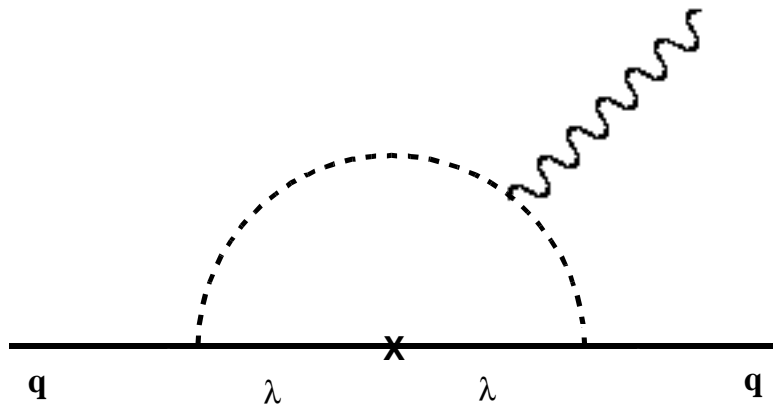
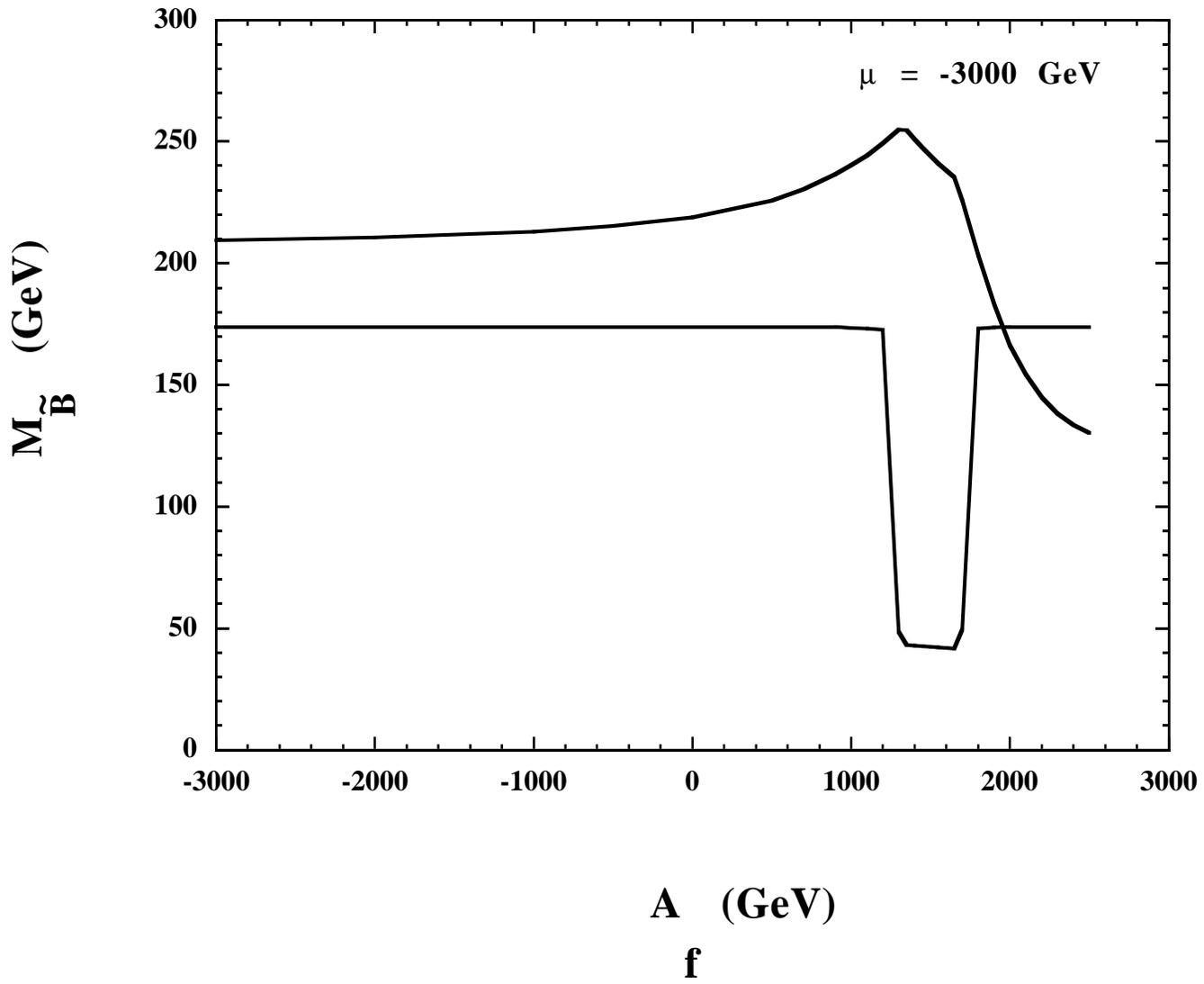
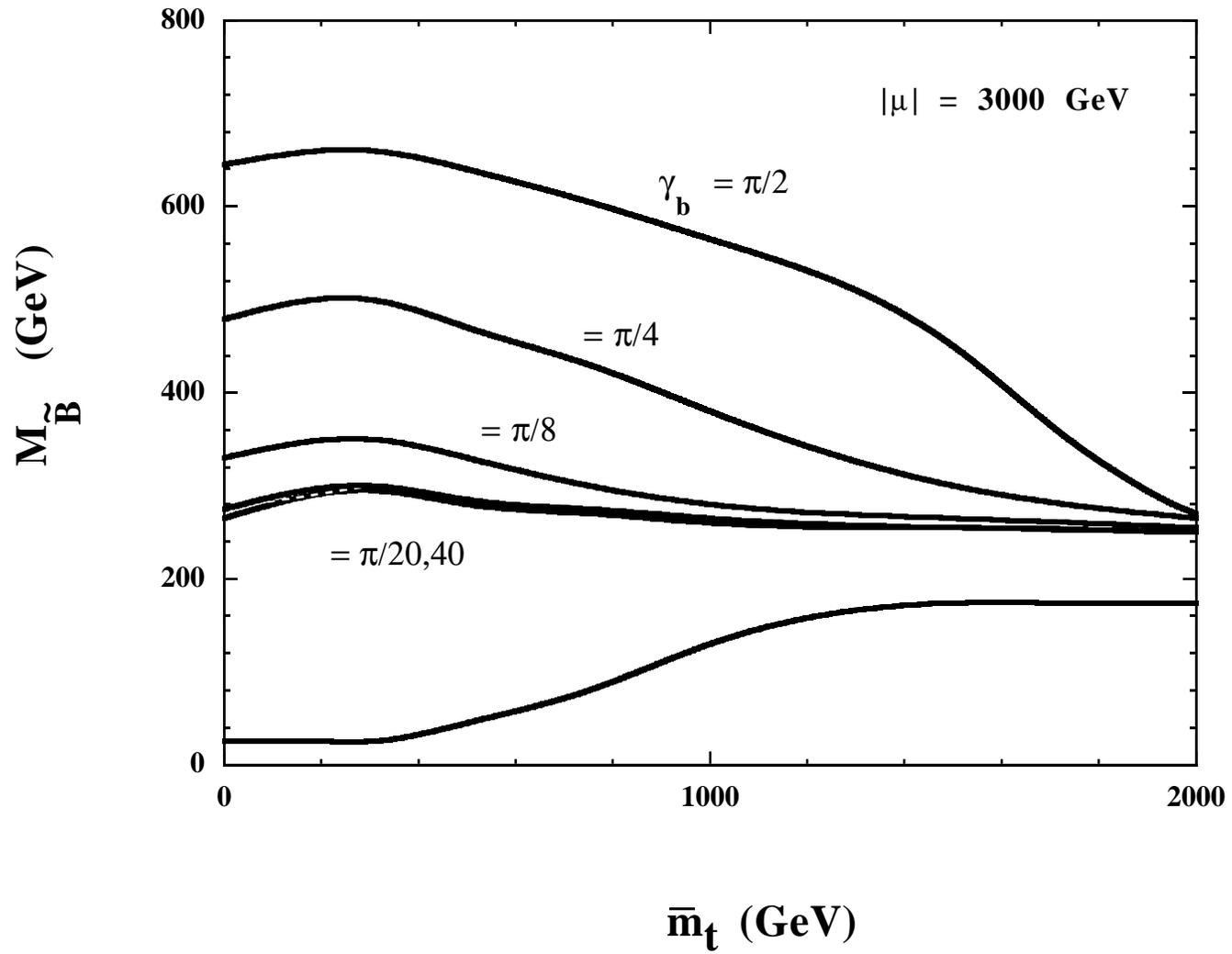
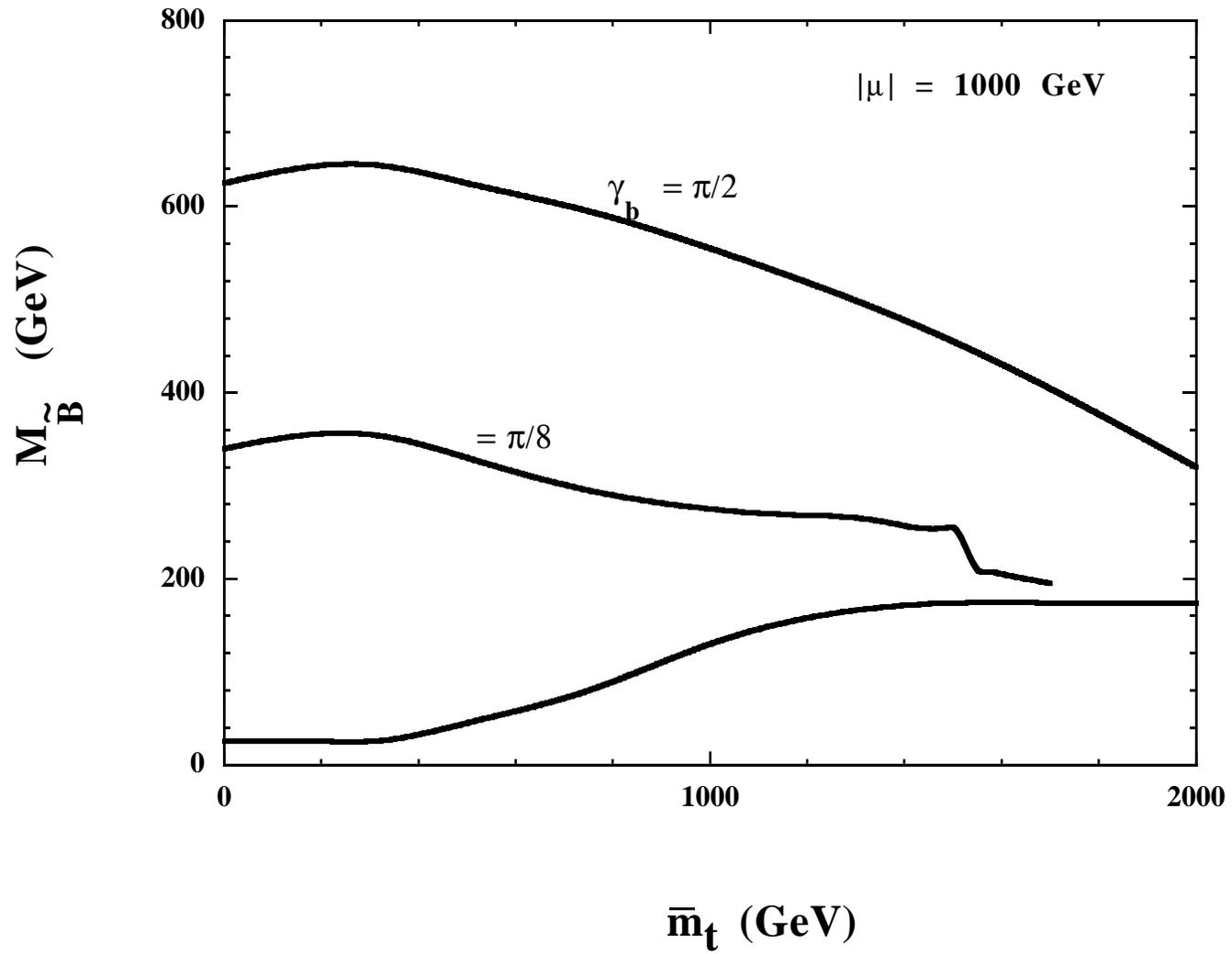
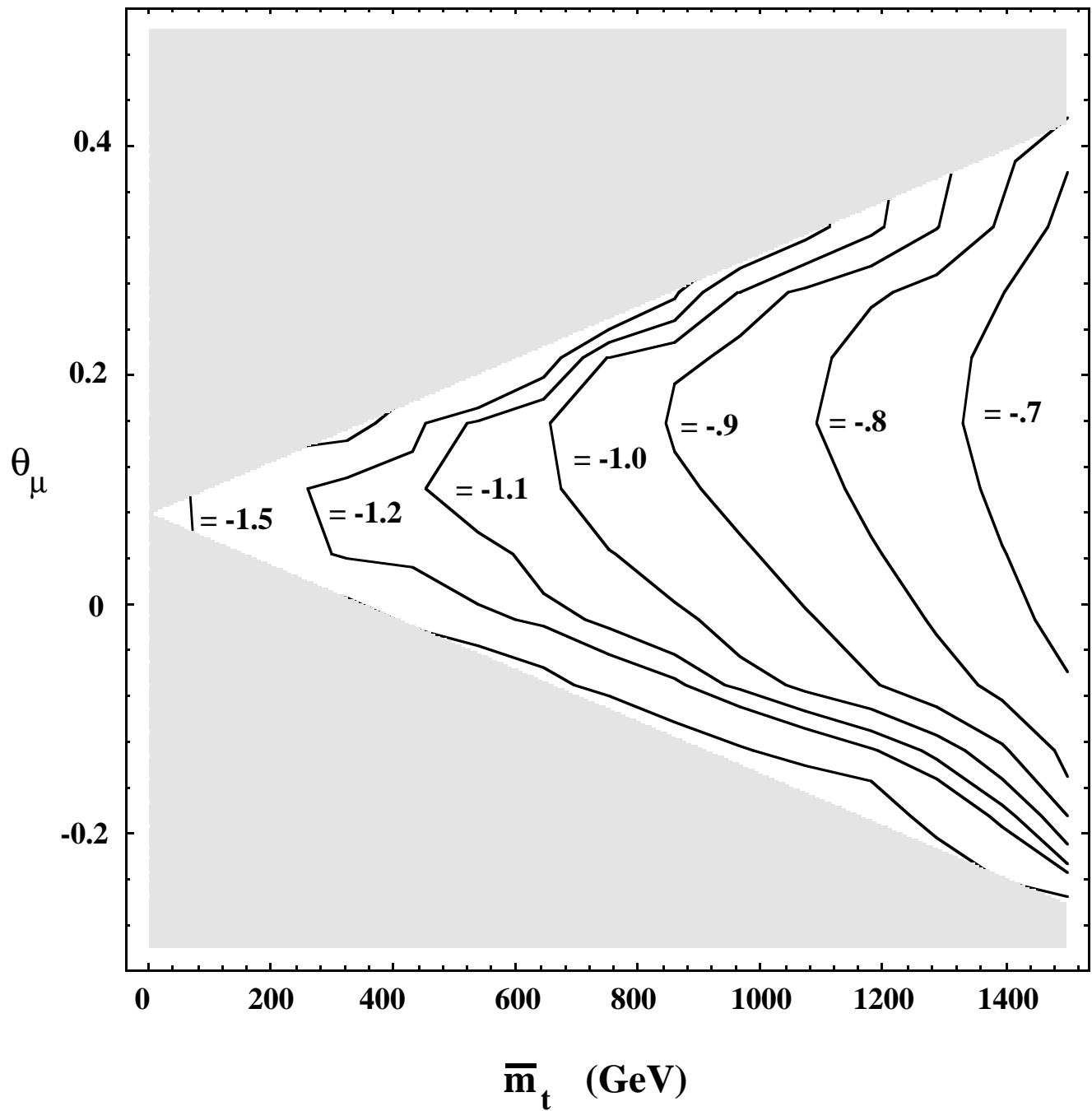
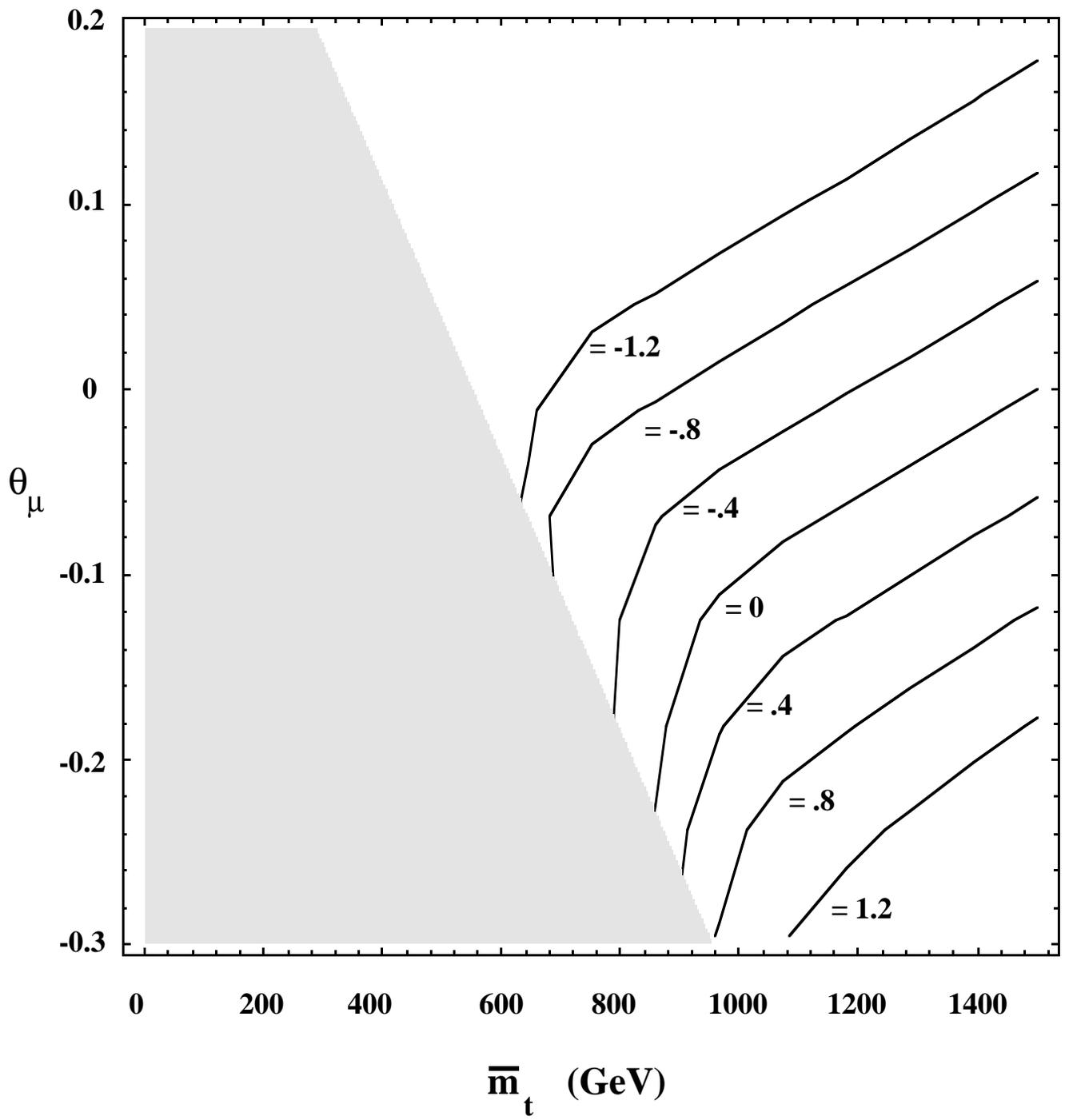
\end{document}